\documentclass[twocolumn,showpacs,preprintnumbers,amsmath,amssymb]{revtex4}

\usepackage{graphicx}
\usepackage{subfig}
\usepackage{dcolumn}
\usepackage{bm}

\usepackage{color}

\newcommand{\beq}[1]{  \begin{equation} \label{#1} }  
\newcommand{\eeq}{     \end{equation}}  
\newcommand{\bal}[1]{\begin{align} \label{#1} }

\newcommand{\rf}[1]{(\ref{#1})}

\def\bd#1{\mbox{\boldmath$\displaystyle\mathbf{#1}$} }
\def\dd{\operatorname{d}} 
\def\grad{\operatorname{grad}} 
\def\div{\operatorname{div}} 
\def\om{k}

\begin{document} 

\preprint{APS/123-QED}

\title{Acoustic cloaking in 2D and 3D 
using finite mass}  

\author{ Andrew N. Norris} 

  \email{norris@rutgers.edu}
\affiliation{Mechanical and Aerospace Engineering, Rutgers University, Piscataway NJ 08854}
\homepage{http://www.mechanical.rutgers.edu/norris}
\date{\today}

\begin{abstract}

Fundamental features  of rotationally symmetric  acoustic cloaks with anisotropic inertia are derived. Two universal relations are found to connect the radial and transverse phase speeds and the bulk modulus  in the cloak. 
 Perfect cloaking occurs only if the radial component of the  density  becomes infinite at the cloak inner boundary,   requiring an infinitely massive cloak.   A practical cloak of  finite mass is defined in terms of its  effective visible radius, which  vanishes  for perfect cloaking.  Significant cloaking is obtained when  the effective visible radius  is subwavelength, reducing the total scattering cross section, and may be achieved even as  the interior radius of the cloak is large relative to the wavelength.  Both 2D vs. 3D effects are compared as we illustrate how the spatial dependence of the cloaking parameters effect the total cross section.  


\end{abstract}


\maketitle


The fundamental   
 observations of Pendry et al. \cite{Pendry06} and of Leonhardt \cite{Leonhardt06} that the electromagnetic equations remain invariant under spatial transformations has generated significant   interest in the possibility of passive  acoustic cloaking.   The idea is to map a region around a single point in such a way that the mapping is one-to-one everywhere except at the point, which is mapped into the cloak inner boundary. 

Consequences of the  transformation method for EM cloaking have been  studied extensively. Thus,   Rahm  et al. \cite{Rahm07}  demonstrate   square cloaks; Chen and Chan  \cite{Chen07b} provide an interesting mapping that rotates the EM field inside the cloaking region;    Chen et al. \cite{Chen07a} examine in detail the interaction with a spherical cloak characterized by the linear radial transformation \cite{Pendry06};   Zhang et al. \cite{Zhang08} consider  cloaking in inhomogeneous environments;    Huang et al. \cite{Huang07}  describe a method to realize   cylindrical cloaking  using a  structure  with  ``normal" EM layers;  Kwon and Werner  \cite{Kwon08} extend the linear transformation to an eccentric elliptic annular geometry lacking rotational symmetry; a quadratic cloak was proposed by 
and investigated by Yan et al. \cite{Yan07}; Cai et al. \cite{Cai07} use a  similar quadratic  spatial transformation to obtain a  smooth transition in the 
 moduli at the outer interface; and  Leonhardt and Philbin \cite{Leonhardt06a} provide a general theory of the transformation method in the context of EM waves.  
 
An acoustic fluid is described by a bulk modulus $K$ and density $\rho$, but cloaking cannot occur if the bulk modulus  and density are  scalar quantities. It is possible to obtain acoustical cloaks by assuming a two parameter density tensor \citep{Cummer07,Chen07,Cummer08}.   A  tensorial density is not ruled out on basic principles \cite{Milton06} and may in fact be realized through so-called metamaterials.  The general context for anisotropic inertia is the   Willis equations of elastodynamics \cite{Milton07} which  
Milton et al. \cite{Milton06} showed  are the natural counterparts to the EM equations that remain invariant  under spatial transformation.   Acoustic cloaking has been demonstrated, theoretically at least, in both 2D and 3D:   a spherically symmetric   cloak was discussed  by Chen and Chan \citep{Chen07} and by Cummer et al. \citep{Cummer08}, while   Cummer and Schurig \cite{Cummer07} described a 2D cylindrically symmetric acoustic cloak.  These papers use a  linear transformation  based on prior EM results    in 2D \cite{Schurig06}.   While the focus here is on passive cloaking, we note that Miller \cite{Miller06} describes possible active acoustic cloaking strategies.  


The 2- and 3D cloaks of  \cite{Cummer07,Chen07,Cummer08} are shown here to be special cases of a more general class of acoustic cloaks.  
We will show that the linear transformation proposed in \cite{Pendry06} and examined by 
 \cite{Chen07a,Cummer08} is just a special case of a general class of transformations possible in acoustic cloaking. 
The arbitrary nature of  the spatial transformation  function has been noted   \cite{Cai07} but not utilized much as yet, except for quadratic cloaks \cite{Cai07,Yan07}.  In this paper we show that 
the cloak density tensor  depends upon the arbitrary spatial mapping function.  Issues that have not been previously considered, such as the total amount of mass required for cloaking,  are addressed.  It turns out that perfect cloaking occurs only if infinite mass is available, regardless of the spatial mapping function.  We take the point of view that achieving infinite mass is impractical, but that  effective cloaking with  finite mass is  still possible if the cloak parameters are appropriately chosen.   We begin with a derivation of the general form of the density tensor and bulk modulus necessary for acoustic cloaking with rotational symmetry in both   2D and 3D.


The  equations governing small amplitude disturbances in an acoustic metamaterial are 
\beq{1}
{\bd \rho} \dot{\bd v} = - \grad p , 
\qquad
\dot{p} = - K \div {\bd v}. 
\eeq
Here, 
$p ({\bd x},t)$ and ${\bd v}({\bd x},t)$ are the pressure  and particle velocity field variables, and $K({\bd x})>0$ is the bulk modulus.  The density, ${\bd \rho}({\bd x})$, is a symmetric second order tensor.    Traditional acoustics 
  corresponds of course to the diagonal form  ${\bd \rho}=\rho {\bd I}$.     
Equations \rf{1} imply that the pressure satisfies the generalized acoustic wave equation 
\beq{2}
K \div  \big(  {\bd \rho}^{-1}\grad p \big) - \ddot{p} =0. 
\eeq
 We  examine   the 2D and 3D configurations in parallel.

Consider spatially inhomogeneous anisotropic materials that are locally transversely isotropic, that is, characterized by an axis of symmetry and an orthogonal hyperplane of isotropy.  The axis of symmetry is in the radial direction, such that the density matrix has the form 
 \beq{4}
{\bd \rho} = \rho_r(r) \hat{\bd x} \otimes \hat{\bd x}  
+ \rho_\perp(r) ({\bd I}- \hat{\bd x} \otimes \hat{\bd x},  
\eeq
where $r = |{\bd x}|$ and  $\hat{\bd x} = {\bd x}/r$.   The bulk modulus, being a scalar, depends only on radius, $K(r)$. 
The governing equation for $p$ is   
\beq{5}
   \frac{K(r)}{r^{d-1}}
\frac{\partial }{\partial r} \bigg( \frac{r^{d-1}}{\rho_r(r) } \frac{\partial p}{\partial r} \bigg)
+ \frac{K (r)  }{ r^2 \rho_\perp(r)} \triangle_\perp p - \ddot{p} = 0, 
\eeq
where $d$ is the dimension (2 or 3) and $\triangle_\perp p$ is the Beltrami-Michel operator.  

The central idea  is to contract or shrink the radial coordinate $r\rightarrow f(r)$, with the pressure parameterized in terms of the contracted radial coordinate as 
$
p ({\bd x},t) = P\big( f(r),\hat{\bd x} , t\big)$. 
The scalar governing equation \rf{5} becomes
\beq{6}
f'(r)   \frac{ K(r)}{r^{d-1} }
\frac{\partial }{\partial f} \bigg( \frac{r^{d-1} f'}{\rho_r(r) } \frac{\partial P}{\partial f} \bigg)
 +   \frac{K  (r)  }{  r^{2} \rho_\perp } \triangle_\perp P - 
 \ddot{P}= 0, 
\eeq
where $f' = \dd f/\dd r$. 
At the same time, the  scalar wave   equation is $\nabla^2_f P   -\ddot{P}= 0$
in  the  coordinates $(f, \hat{\bd x} )$ 
where $\nabla^2_f$ is 
the Laplacian in these  un-contracted coordinates.  The pressure simultaneously satisfies this uniform  wave equation in $(f(r),\hat{\bd x})$ and eq.  \rf{6} in $(r,\hat{\bd x})$
 if and only if the following three conditions are met: 
\beq{10}
\rho_r   = \frac{ r^{d-1}}{ f^{d-1}} f' , 
 \qquad
 {\rho_\perp} = \frac{ r^{d-3}}{ f^{d-3} f'},
 \qquad
K = \frac{ r^{d-1}}{ f^{d-1} f'} .
\eeq

\begin{figure}[b]
				    \includegraphics[width=2.5in , height=2in 					]{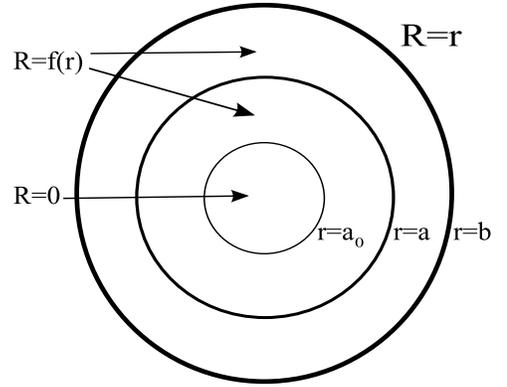} 
\caption{The cloak occupies $a< r<b$ with uniform properties outside and the "`hidden"' object inside.  The cloak properties depend upon a monotone contraction mapping $f(r)$ within the cloak, such that 
	$f(b)=b$, $f(a)<a$.  It is useful to think of an invisibility boundary $r=a_0$ inside the cloak, where $f(a_0)=0$.  The pressure satisfies a uniform wave equation in the mapped radial variable $R(r)$.}
		\label{f1}   
	\end{figure}

The exterior of $r=b$ is a uniform and isotropic acoustic medium with 
$K=1$, ${\bd \rho} = {\bd I}$.  Continuity at $r=b$ requires matching of the pressure and the normal (radial) velocity, and hence the radial acceleration.  The latter 
follows from eq. \rf{1} 
as $\dot{v}_r = -\rho_r^{-1} \partial p/\partial r$, or  $\dot{v}_r  = -  (f/r)f^{d-1}   \partial P /\partial f  $. 
Hence, all that is required for continuity at $r=b$ is that   $f$ be continuous across the boundary, which is accomplished by requiring $f(b) = b$.  
Surprisingly, the two conditions at $r=b$ reduce to this single requirement.  



 Equations \rf{10} provide  relations for the anisotropic acoustic material properties for a given contraction   $f(r)$.  
 It is useful to express them  in terms of the radial and azimuthal phase speeds: 
$
c_r  =  \sqrt{  {K}/{\rho_r} }$ and 
 $c_\perp  =  \sqrt{ {K}/{\rho_\perp} }$. 
The mass density tensor can then be expressed  
$
{\bd \rho} =  K  
\big(  c_r^{-2} {\bd I}_r  +  c_\perp^{-2}{\bd I}_\perp \big) $. 
Equations  \rf{10}  imply the identity 
\beq{11}
K^{d-2}    =  \rho_r {\rho_\perp}^{d-1},
\eeq
independent of the choice of contraction function $f(r)$.    
The quantity $K \rho_r $ is the square of the radial acoustic impedance, $z_r \equiv \sqrt{K \rho_r}$.  
Equation  \rf{11} implies that the identity $z_r = c_\perp^{d-1}$  is required for cloaking. 
 The three equations   \rf{10} can be replaced by \rf{11} along with relations for the phase speeds in terms of $f$: 
\beq{110}
c_r   =    {1}/{f'} , 
 \qquad
c_\perp =      {r}/{f}  . 
\eeq
Note that $f'$ is required to be positive. 
 The original quantities can be expressed in terms of the phase speeds as 
 \beq{-3}
 \rho_r =  c_r^{-1}{c_\perp^{d-1}},\qquad
   \rho_\perp =c_r c_\perp^{d-3}  ,
   \qquad   
   K = c_r c_\perp^{d-1} .
 \eeq
One  could, for instance, eliminate $f$ as the fundamental variable defining the cloak in favor of  $c_\perp (r)$, from which all other quantities can be determined from the differential equation  relating the speeds: 
$
\big(  {r}/{  c_\perp}\big)' = 1/c_r $. 
We assume the cloak occupies $\Omega = \{0< a\le r \le b\}$ with uniform  acoustical  properties   $K=1$, ${\bd \rho} = {\bd I}$ in the exterior. 
Then  $c_\perp(r)$ follows  by integration, 
\beq{90}
\frac{1}{c_\perp } = \frac{b}{r} -  \frac{1}{r}\int_r^b \frac{\dd r}{c_r},
\qquad a\le r \le b.  
\eeq

Can  the cloak  density be isotropic?  This will occur  if $c_r = c_\perp$, 
 which requires that $f' =   f/r$, implying 
$f = \gamma r$ with  
  $\gamma$ constant.  Impedance matching at the cloak outer boundary $r=b$  requires  $f(b)=b$.   This can only be satisfied in the trivial case of  $\gamma  = 1$, implying $f\equiv r$.   While this eliminates the possibility of isotropic cloaks, it is perhaps a satisfying outcome.  
  
  
\begin{figure}[htbp]
 \subfloat[2D, $\alpha = 1$, $\sigma_{tot}=1.18$] {\label{f2-a}\includegraphics[width=0.22\textwidth]{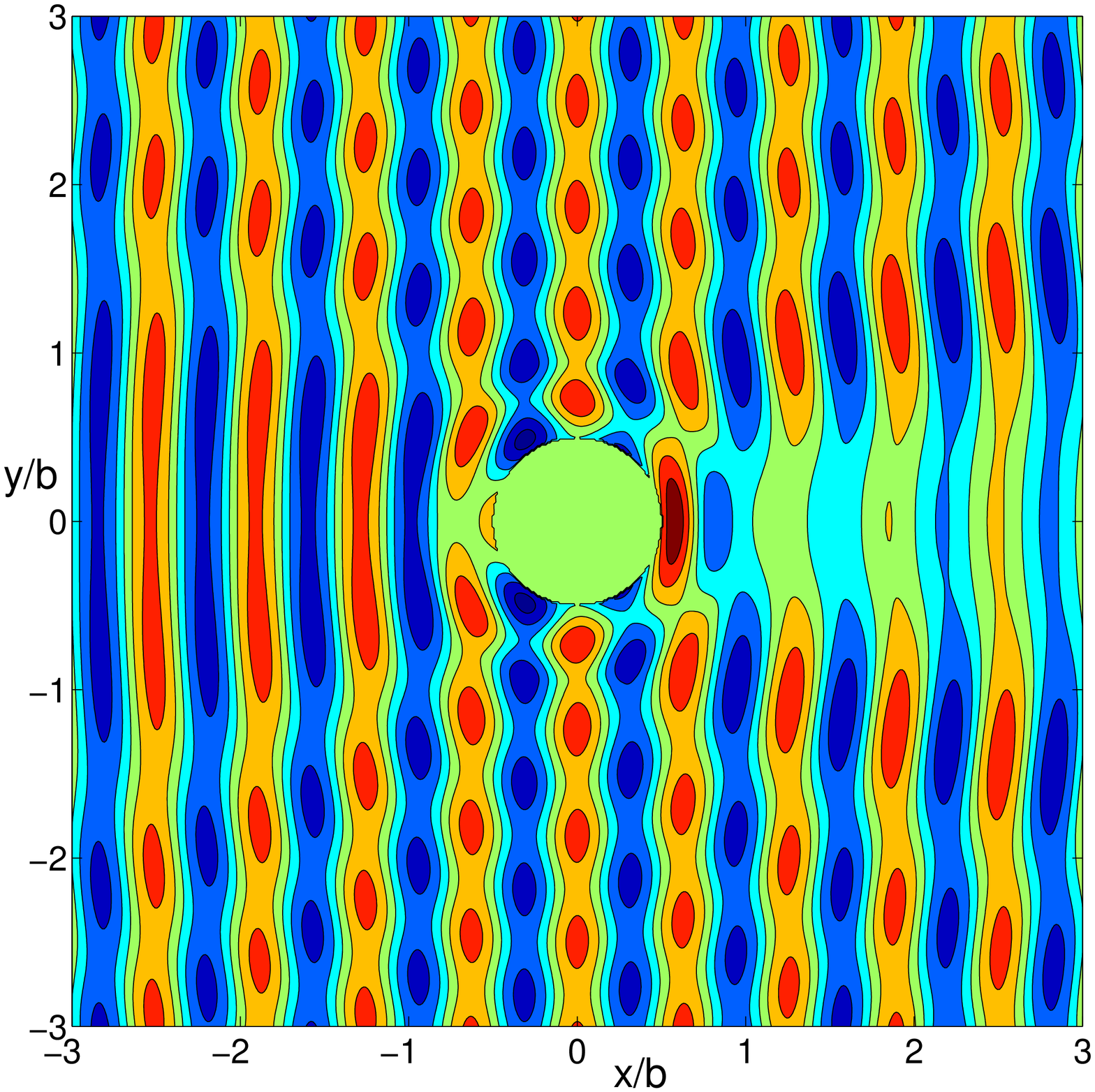}}    \subfloat[3D, $\alpha = 1$, $\sigma_{tot}= 0.49$] {\label{f2-b}\includegraphics[width=0.22\textwidth]{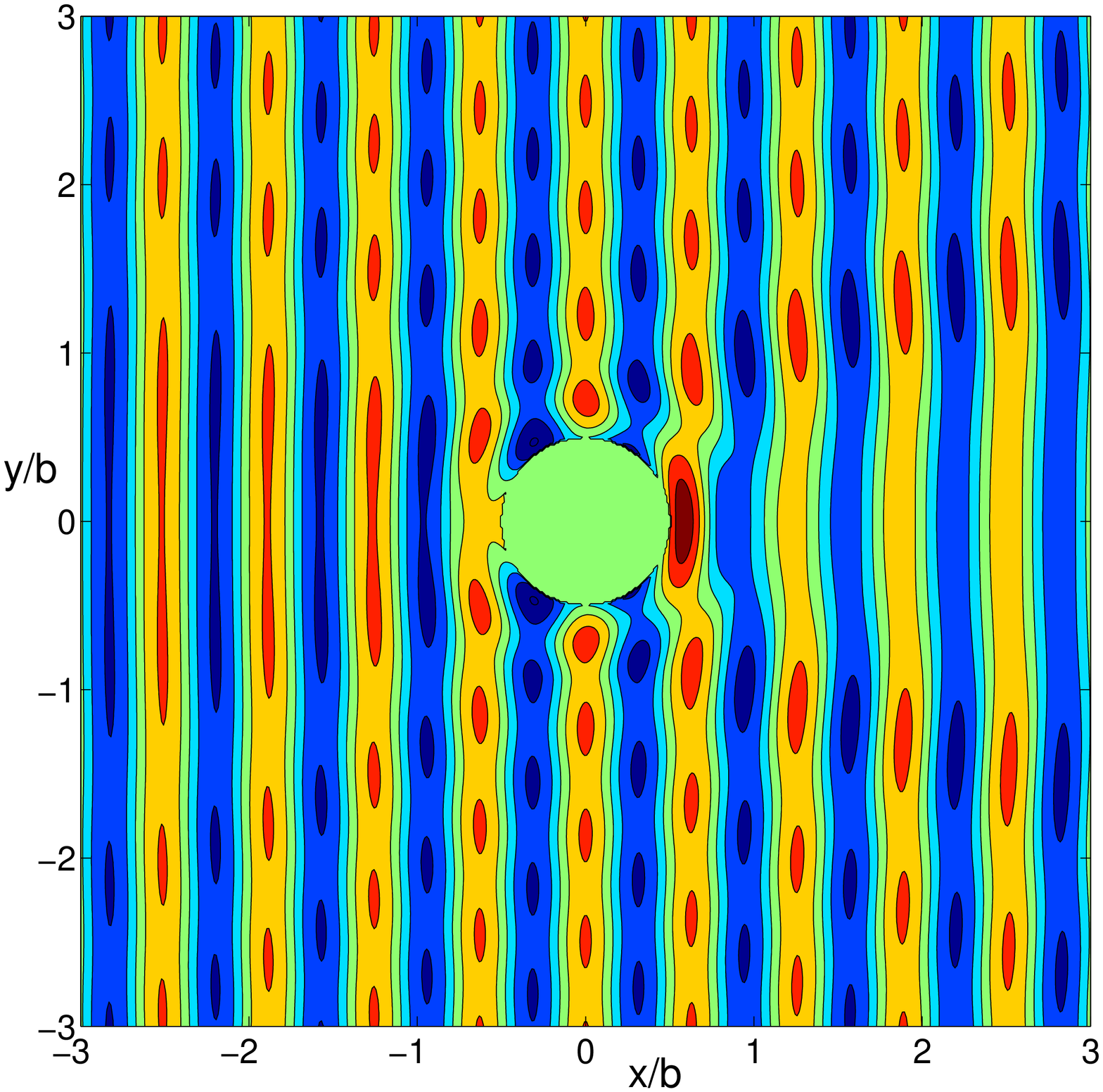}}  \\ \vspace{-.05in} \subfloat[2D, $\alpha = 4$, $\sigma_{tot}=0.06 $] {\label{f2-c}\includegraphics[width=0.22\textwidth]{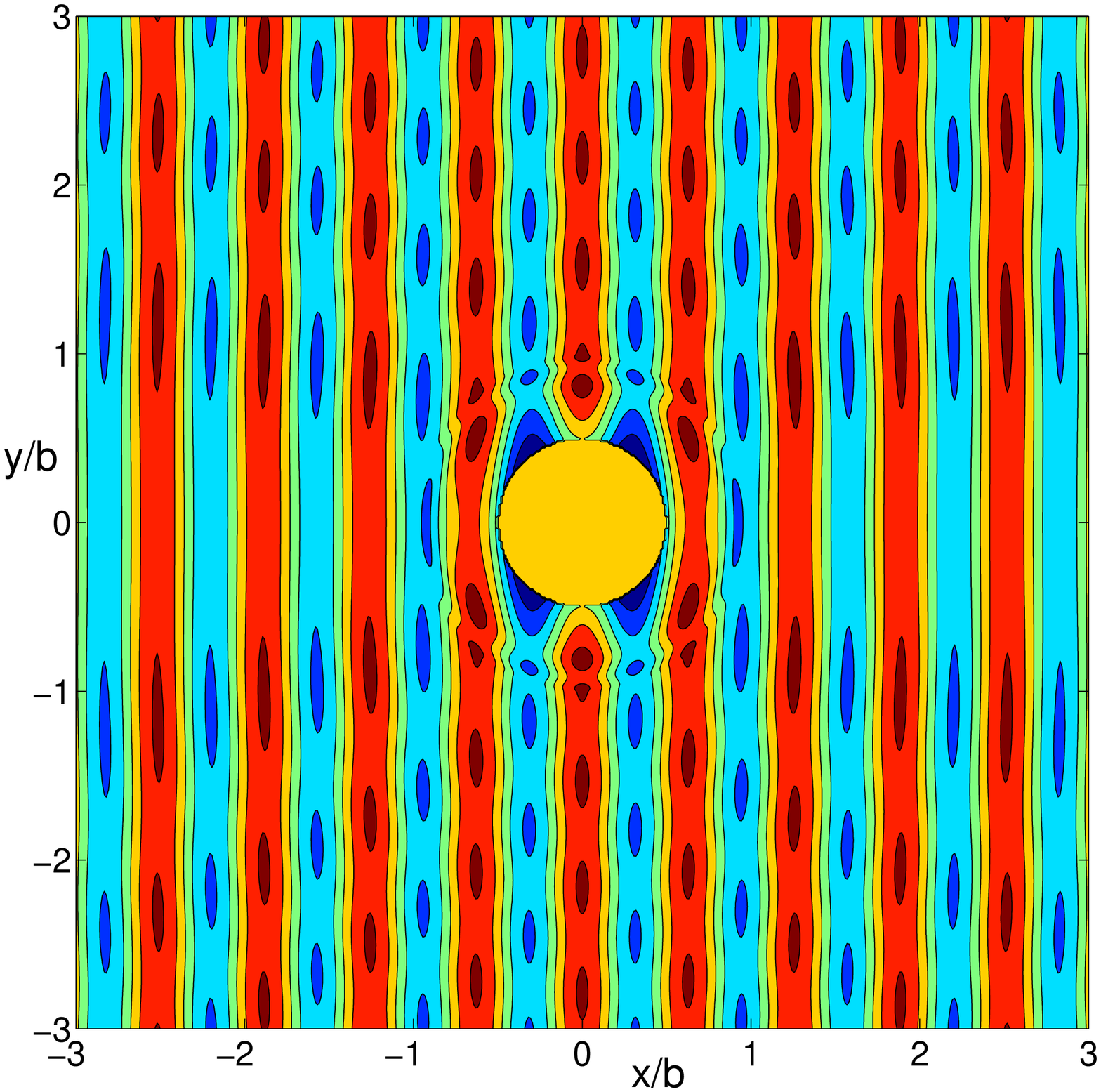}}   \subfloat[3D, $\alpha = 4$, $\sigma_{tot}=  10^{-4}$] {\label{f2-d}\includegraphics[width=0.22\textwidth]{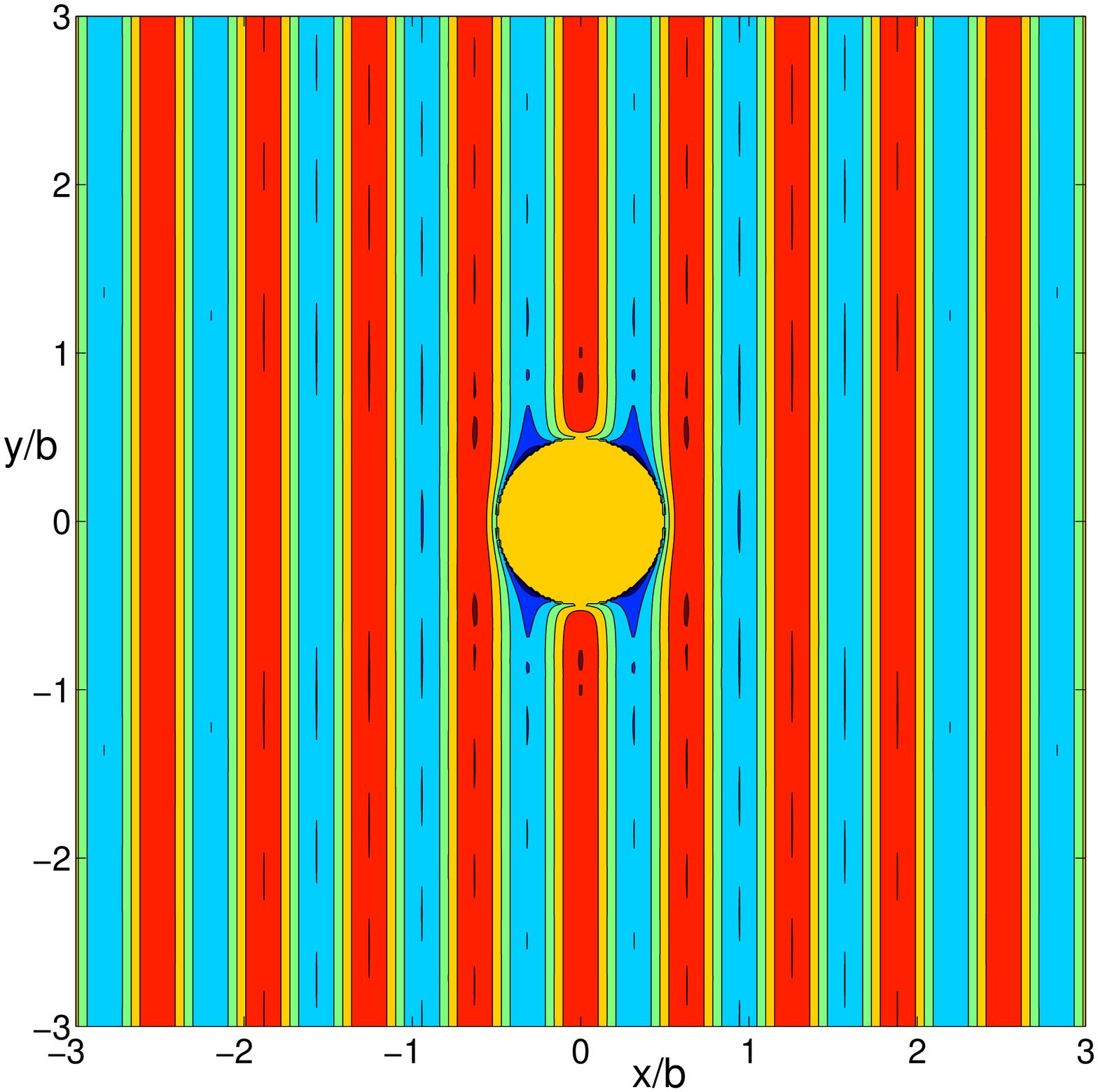}} 
 	\caption{A  plane wave is incident from the left with frequency  
$\om = 10$ on the cloak defined by eq. \rf{21} with $(b,a,a_0)= (1,0.5, 0.35)$.  The exponent is
$\alpha = 1$ in (a) and (b)  implying a virtual inner radius of $ f(a) = 0.23$.  For (c) and (d), 
$\alpha = 4$ giving $f(a)=0.003$.
 } 
		\label{f2}  
\end{figure}

Perfect cloaking requires that $f$ vanish for some $r= a_0>0$.  It is clear that $c_\perp$ and $z_r$ blow up as 
$r\downarrow a_0$.  Hence the product $K \rho_r$ blows up, but do $K $ and $\rho_r$ each become unbounded?  Consider $f \propto (r-a_0)^\alpha $ near $a_0$ for  $\alpha $ constant and non-negative.  No value  of $\alpha >0$ will keep the radial density $\rho_r$ bounded, but the unique choice $\alpha = 1/d$ ensures that the  bulk modulus $K (a_0)$ remains finite.    We will see below that this power law is not a practical choice for the cloak mapping function.
Note that the azimuthal density $\rho_\perp$ has a finite limit in 2D for power law decay 
$f \propto (r-a_0)^\alpha $, while $\rho_\perp$ remains finite in 3D  iff $\alpha\le 1$, otherwise it blows up.   Similarly, the radial phase speed scales as $c_r \propto (r-a_0)^{1-\alpha} $, which remains finite  for $\alpha\le 1$,  blowing up otherwise.
These results are summarized in Table I. 

 \begin{table}\label{tab1}
\caption{Behavior of quantities near the vanishing point $r=a_0$ for the scaling
$f \propto x^\alpha $ as $x= r-a_0 \downarrow 0$.  The total radial mass $m_r$ is defined in eq. \rf{22}. 
}
\begin{center}
\begin{tabular}{|c|c|c|c|c|c|c|}
\hline  & & & & & & \\  
\, dim \,  & $\rho_r$ & $\rho_\perp$ &  $c_r$ & $c_\perp$ & $K$ & $m_r$ 
 \\  
 & & & & & & \\  
 \hline & & & & & &  \\ 
2 &   $x^{-1}$ & $x$ & $x^{1-\alpha}$ & $x^{-\alpha}$ & $x^{1-2\alpha}$ & $\ln x$ 
 \\ 
 & & & & & & \\  
  \hline & & & & & & \\ 
3 &  \, $x^{-1-\alpha}$\, &\, $x^{1-\alpha}$\, & \,$x^{1-\alpha}$ \,&\, $x^{-\alpha}$ \,&\, $x^{1-3\alpha}$\,  &\, $x^{-\alpha}$ \,
 \\ 
 & & & & & & \\  
 \hline
 \end{tabular}
\end{center}
\end{table}


We introduce a nondimensional measure of the total amount of cloaking mass present in the radial component of the inertia tensor: 
\beq {22}
m_r \equiv   \frac{ \int_\Omega \dd V \rho_r }{ \int_\Omega \dd V }
= \frac{d}{b^d-a^d}\int\limits_a^b\dd r \, r^{2(d-1)}
\frac{ f'(r) }{f^{d-1} }. 
\eeq
Explicitly, 
\beq{43}
m_r 
=
\begin{cases} 
\frac{2}{b^2-a^2}\big[ 
b^2 \ln f(b) - a^2 \ln f(a) - 2 \int_a^b \dd r \, r  \ln f(r)
\big] 
, 
\\
 & \\ 
\frac{3}{b^3-a^3}\big[ 
\frac{a^4}{f(a) } - \frac{b^4}{f(b) } + 4\int_a^b \dd r \, \frac{r^3 }{f(r)}
\big]  ,  
\end{cases}
\eeq
in 2D and 3D, respectively. These forms indicate not only that $m_r \rightarrow \infty$ as $f(a) \rightarrow 0$, but also the form of the blow-up.  To leading order, 
$m_r = \frac{2a^2}{b^2-a^2}\ln \frac{1}{f(a)} + \ldots$ in 2D and 
$m_r = \frac{3a^3}{b^3-a^3} \frac{a}{f(a)} + \ldots$ in 3D.  The blow-up of $m_r$ occurs no matter how $f$ tends to zero; that is, the infinite mass is an unavoidable  singularity. 

The idea behind the contraction function $f(r)$ is to map the entire interior of $r\le b$ into an annular subset: $\Omega$.   This in turn implies that perfect cloaking requires infinite mass $m_r$.  We take the point of view that $m_r$ must remain finite, which can be achieved by making $f>0$ in $\Omega$.  To be specific, consider the analytic continuation of  $f(r)$ into the ``cloak'' $\{r:  0\le r < a\}$, 
where we allow  $f(r)$ to vanish.   
Consider the power law form for the mapping function, $f(r) = f^{\alpha} (r) $ where 
\beq{21}
f^{(\alpha)}(r) = b\big( \frac{r-a_0}{b-a_0}\big)^\alpha, 
\eeq
for $0\le a_0< a$.  
This yields    finite mass $m_r$, but it blows up as $a \downarrow a_0$ for every $\alpha >0$. 

It follows from \rf{43} that the dependence of the total mass on $\alpha$ is linear  in 2D: $m_r^{(\alpha)} = \alpha m_r^{(1)}$. 
  In other words, in 2D the mass required is proportional to the log of the effective vanishing radius. 
  For 3D, the leading order term in eq. \rf{43} implies 
 $
 m_r^{(\alpha)} \approx \beta^{-1}   \big( \beta  m_r^{(1)}  \big)^\alpha $, 
  where $\beta = \frac13 \frac{b}{a}( \frac{b^3}{a^3} -1)$.   The total mass grows exponentially with $\alpha$, all other parameters being fixed.  
However, we can still achieve effective cloaking with finite $m_r$.  The idea is to make $f(a)$ subwavelength.   This does \emph{not} require that $ a$ be subwavelength.

We assume time harmonic motion, with the factor $e^{-i \om t}$ understood but omitted.   Consider a cloak of finite mass with zero pressure on the interior surface $r=a$.  
The total response for plane wave incidence is 
 \begin{widetext}
\beq{41}
p = p_0 e^{i \om R(r) \cos\theta } 
- p_0 \sum\limits_{n=0}^\infty  
\begin{cases} i^n (2-\delta_{n0})
J_n( \om R(a) ) 
\frac{ H^{(1)}_n (\om R(r) )} { H^{(1)}_n (\om R(a) )}
\cos n \theta  , & 2D,
\\
& \\
i^n (2n+1)  j_n( \om R(a) ) 
\frac{ h^{(1)}_n (\om R(r) )} { h^{(1)}_n (\om R(a) )}
P_n ( \cos  \theta )  , & 3D,
\end{cases}
\qquad
R(r) = \begin{cases} f(r), & a\le r< b,
\\
& \\
r, & b\le r < \infty . 
\end{cases}
\eeq
\end{widetext}

The total scattering cross-section,  and hence the total energy scattered, depends on the forward scattering amplitude through the optical theorem: $\sigma_{tot} = 4\pi \om^{-1} \text{Im}g(\hat{\bd e}_z)$, 
where $g$ defines the leading order far-field $p = p_0 e^{i \om z} + p_0 g(\hat{\bd x}) r^{-(d-1)/2} (i 2 \pi/\om)^{(3-d)/2}e^{i\om r} $. 
The scattering cross-section is easily computed from \rf{41}, and is   
 dominated in the small $\om R(a)$ limit by the $n=0$ term, with leading order approximations 
\beq{52}
\sigma_{tot} = 
\begin{cases}
\frac{\pi^2}{\om} |\ln \om f(a)|^{-2} + \ldots , & 2D,
\\
& \\
4\pi f^2(a) + \ldots , & 3D.
\end{cases}
\eeq
The faster decay of $\sigma_{tot}$  explains the greater efficacy of the finite mass cloak in 3D.  Considered as a function of the power $\alpha$,  in 3D  the mass $m_r$ \emph{increases} exponentially while the cross-section $\sigma_{tot}$  \emph{decreases} exponentially.  In  2D the rates of increase  and decrease of   $m_r$ and $\sigma_{tot}$ are algebraic (linear) or less.   The comparison suggests that cylindrical cloaks require that the visibility radius $f(a)$ be very small.  Figure 2 illustrates clearly the disparity between chylindrical and spherical cloaking.  Thus,  for $f(a)/b =0.003$ the 3D cross-section is neglible   (Fig.  2.d) but for 2D the cross-section is two orders of magnitude larger (Fig.  2.c).   Note that Ruan et al. \cite{Ruan07} found that  the perfect cylindrical EM cloak is sensitive to perturbation.  This sensitivity is  evident  from the present analysis through the dependence on the   length  $a-a_0>0$ which measures the departure from perfect cloaking $(a=a_0)$.  

In  conclusion, we have shown how to generate acoustic cloaks in 2D and 3D using a general  radial transrformation $r\rightarrow f(r)$. Regardless of the form of $f$, two universal equations relate the cloak parameters, e.g. 
$ K = c_r c_\perp^{d-1}$ and $\big(  {r}/{  c_\perp}\big)' = 1/c_r $.   Perfect cloaking requires not only that the density  $\rho_r \rightarrow \infty $ at the inner radius, but the associated  total mass $m_r$ is also infinite.  This can be avoided in practice since the scattering cross-section $\sigma_{tot}$ can be made as small as desired by  selecting the value of  $f(a)\ll a$ so that   
 $\om f(a)\ll 1$ even as  $\om a =$O$(1)$.   We  showed that the form of the cloak mapping function $f(r)$ is important and the effect of the power $\alpha$ is markedly  different in 2D vs. 3D.
The feasibility of   cloaking depends on the development of real  metamaterials with the desired unusual physical properties.  This is an active research area  \cite{Li04a,Fang06,Milton07b} with tremendous potential for future applications.





\end{document}